%% file: ieee-main.tex
\documentclass[conference]{IEEEtran}
\usepackage[caption=false]{subfig}
\usepackage{cite}
\usepackage{amsmath,amssymb,amsfonts}
\usepackage{algorithmic}
\usepackage{algorithm}
\usepackage{graphicx}
\usepackage{textcomp}
\usepackage{xcolor}
\usepackage{multirow}
\usepackage{hyperref}
\def\BibTeX{{\rm B\kern-.05em{\sc i\kern-.025em b}\kern-.08em
    T\kern-.1667em\lower.7ex\hbox{E}\kern-.125emX}}
\begin{document}

\newcommand{\redtext}[1]{\textcolor{black}{#1}}

\newif\ifcomment
\commenttrue

\newcommand{\yf}[1]{\ifcomment{\color{blue}
\emph{[FEI: #1]}}\fi}
\newcommand{\ding}[1]{\ifcomment{\color{green}
\emph{[AD: #1]}}\fi}

\title{USBSnoop - Revealing Device Activities via USB Congestions}

\author{
\IEEEauthorblockN{Davis Ranney\IEEEauthorrefmark{1}, Yufei Wang\IEEEauthorrefmark{1}, A. Adam Ding\IEEEauthorrefmark{2}, Dr. Yunsi Fei\IEEEauthorrefmark{1}}
\IEEEauthorblockA{\IEEEauthorrefmark{1}Electrical and Computer Engineering, Northeastern University, Boston, USA \\
\IEEEauthorrefmark{2}Mathematics, Northeastern University, Boston, USA \\
\{ranney.d, wang.yufei1, a.ding, y.fei\}@northeastern.edu}
}


\maketitle

\begin{abstract}
The USB protocol has become a ubiquitous standard for connecting peripherals to computers, making its security a critical concern. 
A recent research study demonstrated the potential to exploit weaknesses in well-established protocols, such as PCIe, and created a side-channel for leaking sensitive information by leveraging congestion within shared interfaces. 
Drawing inspiration from that, this project introduces an innovative approach to USB side-channel attacks via congestion. 
We evaluated the susceptibility of USB devices and hubs to remote profiling and side-channel attacks, identified potential weaknesses within the USB standard, and highlighted the critical need for heightened security and privacy in USB technology. 
Our findings discover vulnerabilities within the USB standard, which are difficult to effectively mitigate and underscore the need for enhanced security measures to protect user privacy in an era increasingly dependent on USB-connected devices.
\end{abstract}


\section{Introduction}
\label{sec:introduction}

The Universal Serial Bus (USB) protocol is now an aptly named nearly universal standard that supports a plethora of peripheral devices to interact with computers. 
Initially designed to simplify external devices' connection to a host with a standard connector and protocol, USB has evolved significantly in functionality and speed. 
USB operates as a packet-based communication system, utilizing the concept of pipes and communication layers, which logically resembles IP-based networking and Peripheral Component Interconnect Express (PCIe). 
A critical aspect of USB technology is the integration of hubs, which serve as instrumental components in expanding the connectivity of a single computer or controller to multiple devices. 
Hubs not only enhance the versatility of USB but also add layers of complexity to the data traffic management~\cite{noauthor_usb_nodate-3}.

Despite USB's widespread adoption and technological advancements, security has not been a primary consideration in its specifications and revisions. 
Historically, most USB-related security concerns necessitated either specifically vulnerable hardware or direct physical access to the device(s), resulting in the underestimation of potential security risks in USB implementations by the USB Implementor's Forum~\cite{tian_sok_2018}. 
For instance, some USB attacks spoof a keyboard and launch commands to download malware~\cite{nohl_badusb_2014}. Others target the human user via phishing or social engineering~\cite{tian_sok_2018}, or even physically destroy the port or device by creating a short circuit in the device~\cite{tian_sok_2018}. 

In parallel to limited USB security concerns, research has been conducted on other related protocols, notably PCIe security.
A pertinent example is the \textit{Invisible Probe} study~\cite{tan_invisible_2021}, which highlights a vulnerability in PCIe's shared bus architecture, demonstrating how devices sharing PCIe lanes can leak their activities to others. 
The principle of the attacks lies in data congestion on PCIe switches among multiple devices, where the adversary utilizes one device as a spy and passively collects its timing/usage information to infer activities on the victim device. 
One attack scenario has a Non-Volatile Memory Express (NVMe) solid-state drive (SSD) acting as a covert probe to monitor a network interface card's (NIC) web traffic. 
The other attack scenario has a specialized NIC monitoring data traffic to a GPU to measure when screen refreshes occur to recover keystrokes and machine learning model architectures~\cite{tan_invisible_2021}.

This research explores a critical yet under-examined aspect of USB security and privacy, demonstrating how USB hubs can be exploited to recover sensitive user data by monitoring patterns of congestion among devices connected to USB hubs. 
Unlike traditional eavesdropping attacks, this method does not require physical access to the device but operates within common configurations used by regular users without needing administrator privileges. 
Using these hub-level side-channels, an attacker can recover a victim user's internet browsing history, keystrokes, and possibly even more sensitive data transmitted across USB.

We first discuss relevant parts of the USB specification and the related attack in Section \ref{sec:background}. 
We then elaborate on detailed methodologies of our approach and its real-world threat models in Section \ref{sec:approach}. 
We present experimental results of two attack scenarios and analysis in Sections \ref{sec:results1} and \ref{sec:results2}. 
We discuss mitigation and improvements that could be made to the USB standard in Section \ref{sec:mitigation}, followed by potential future directions and applications of this research in Section \ref{sec:extensions}.

\section{Background}
\label{sec:background}

\subsection{USB Specification}
\label{sec:seclandscape}
There have been four major revisions of USB specifications. 
USB 2.0 (High Speed), introduced in April 2000, significantly enhanced its predecessor, USB 1.1, by offering data transfer speeds up to 480 Mbps.
USB 2.0 remains the baseline standard in most computer systems today. 
USB 3.X and USB4 further increased data transfer rates to 5 and 40 Gbps, respectively. 
\footnote{USB 3.0, 3.1, 3.2 Gen 1, Superspeed, and SuperSpeed 5 Gbps all describe the same speed of transfers with minor feature differences but are all official marketing names released by USB-IF and are used interchangeably by a lot of device manufacturers. We will use USB 3.X to refer to this speed class as the unique features do not affect the proposed attacks.} 
This paper demonstrates attacks against USB 2.0 and USB 3.X, though our methodology will apply to all current USB revisions.
USB hubs allow controllers and hosts to support multiple devices, and the specification outlines the bandwidth allocation for different devices. 
Most devices do not need to communicate at full speed simultaneously, which is a premise that can be exploited to exfiltrate sensitive information stealthily.

USB is designed as a shared bus connecting a \textit{host} or \textit{root hub} to up to 127 downstream devices across five tiers of external hubs.
At the core of USB functionality is a three-layer communication stack that abstracts the protocol for a more straightforward implementation of data transfer features, including the \textit{function} (application) layer, \textit{device} layer, and \textit{bus interface} layer~\cite{noauthor_usb_nodate-3}, akin to the IP protocol's layers.
The function layer handles device drivers and higher-level software, allowing the operating system to interface with devices easily. 
The device layer manages USB handshakes and protocols via transactions, with hubs routing transaction packets to multiple devices/endpoints. 
The bus interface layer operates differential signaling for upstream and downstream communication. 
Our attack exploits the routing and arbitration of traffic by hubs at the \textbf{device} layer.

For communications between the host and devices, the host initiates all communications with control tokens to indicate when devices can send or will receive data before the data itself~\cite{noauthor_usb_nodate-3}. 
At the device level, data is transmitted in \textit{transactions}, sequences of packets defined by the transaction type. 
USB supports four primary types of data transactions: bulk, isochronous, interrupt, and control, though the attacks in this paper focus specifically on bulk and interrupt transactions. 
\textit{Bulk Transfers} are used for large, non-time-sensitive data transfers, such as file transfers, with error detection and correction mechanisms but no guaranteed timing or bandwidth. 
\textit{Interrupt Transfers} are employed for small, time-sensitive data, typically used by input devices like keyboards and mice.  
This type guarantees regular polling and bandwidth and also error checking.

Data transfers are scheduled based on the transaction type and size, with the host polling each active device for readiness, ensuring efficient communication and preventing collisions. 
Because hosts control communication timing, devices at higher tiers do not necessarily receive additional bandwidth from the root hub. 
USB transactions are organized into \textit{frames} (USB 1.x), which are one millisecond long, and \textit{microframes} (USB 2.0), which are 125-microsecond intervals, to manage bandwidth. 
USB 3.X modifies microframe structure to use \textit{bus intervals} to allow for better bandwidth utilization and reduces reliance on polling, but fundamentally it doesn't affect these attack principles.
Hosts can allow devices to transfer multiple packets per time slot when polled, maximizing bandwidth utilization~\cite{noauthor_usb_nodate-3}.
\autoref{tab:usb2bulk} depicts how bandwidth is utilized for data transfers of a few selected sizes.
As the table shows, multiple bulk transfers can occur per microframe.
How many transfers will be made depends on the size of the data payload of the bulk transfer. 
Data payloads can be any power of 2 between 1 and 512. 
As the data payload increases to a maximum of 512 bytes, fewer transfers are possible per microframe, but more useful data can be sent.
To create a reliable attack, our spy device has to saturate the bandwidth of each time slot in order to measure congestion.

\begin{table}
    \caption{Selected USB 2.0 Bulk Transaction Limits}
    \begin{tabular}{|c|c|c|c|}
    \hline
    \if false 
         Data Payload  & Max Bandwidth  & Microframe Bandwidth & Maximum Transfers  & Bytes per Microframe\\
         (bytes) & (bytes/sec) & per Transfer & per Microframe & of Useful Data\\
    \hline 
    \hline
         1 & 1,064,000 & 1\% & 133 & 133 \\
         \hline
         8 & 7,616,000 & 1\% & 119 & 952 \\
         \hline
         32 & 22,016,000 & 1\% & 86 & 2752 \\
         \hline
         128 & 40,960,000 & 2\% & 40 & 5120 \\
         \hline
         512 & 53,248,000 & 8\% & 13 & 6656 \\
    \fi 
    Data Payload  & Max Bandwidth  &  Maximum Transfers  & Bytes \\
         (bytes) & (bytes/sec) & per Microframe & per Microframe\\
    \hline     
         1 & 1,064,000 & 133 & 133 \\
         \hline
         8 & 7,616,000 &  119 & 952 \\
         \hline
         32 & 22,016,000 &  86 & 2752 \\
         \hline
         128 & 40,960,000 & 40 & 5120 \\
         \hline
         512 & 53,248,000 & 13 & 6656 \\     
    \hline
    \end{tabular}
    \label{tab:usb2bulk}
\end{table}

\begin{figure}
    \centerline{\includegraphics[width=0.9\linewidth]{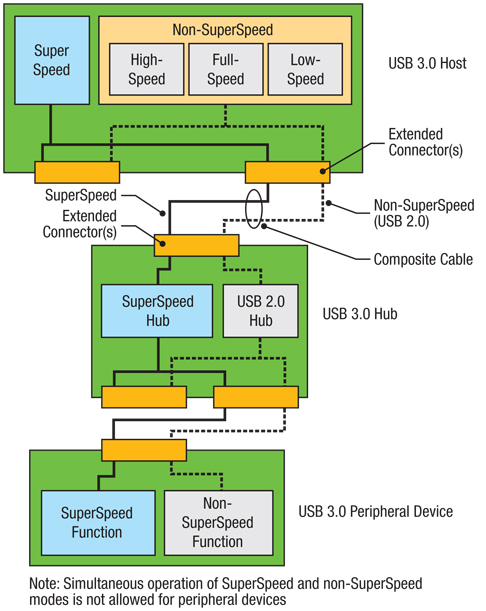}}
    \caption{USB 3.X Hub Logical Architecture, from~\cite{noauthor_usb_nodate-1}}
    \label{fig:usb-3-2-hub}
\end{figure}

USB 3.X connectors and cables are designed with additional wires compared to USB 2.0 to facilitate higher bandwidth without significantly increasing the electromagnetic (EM) signal tolerances and frequency. 
To maintain compatibility with older connectors and devices, when a USB 2.0 device is connected to a USB 3.X port, the traffic from the device does not utilize the additional wires or bandwidth of the USB 3.X standard. 
Instead, it is routed differently within the USB 3.X architecture~\cite{noauthor_usb_nodate-1}, through a dedicated USB 2.0 pathway, shown in \autoref{fig:usb-3-2-hub}.
The newer USB-C connector was released with the USB 3.2 standard but is designed to support USB 2.0 as well as USB4 and newer standards.

USB 2.0 hubs also contain a component called the Transaction Translator (TT)~\cite{atomminer_what_nodate}. The primary role of a TT is to bridge the communication gap between USB 1.1 devices and USB 2.0 devices. 
This is necessary to maintain compatibility with legacy peripherals, keyboards, and mice, which often still use USB 1.1 chips because they do not require the increased bandwidth of USB 2.0 and are cheaper to produce. 
USB 2.0 hubs can have a single TT to handle all downstream ports or multiple TTs dedicated to each downstream port. 
All USB 1.1 traffic is funneled through one translator in USB hubs with a single TT. 
This setup can lead to bottlenecks when multiple Full-speed or Low-speed devices are connected to the hub. 
If several of these devices attempt to communicate simultaneously, they must wait for access to the TT, which handles each request in turn, creating an additional bottleneck in USB communication~\cite{atomminer_what_nodate}.

\subsection{Invisible Probe: Timing Attacks with PCIe Congestion Side-Channel}
\label{sec:iniviprobe}

Invisible Probe~\cite{tan_invisible_2021} explores a security vulnerability in PCIe architecture, which is widely used to connect hardware components like GPUs, SSDs, and network cards to the CPU. 
PCIe can use switches and chipsets to increase system expandability, but one device can intentionally induce congestion on the bus so as to monitor the activities of another device sharing the lanes. 
The prior work designed multiple attacks, with a few that use a specialized Remote Direct Memory Access (RDMA) NIC to spy on a GPU. 
The delay measured by the NIC could be used to infer machine learning model architectures as well as typing patterns accurately when the GPU refreshes the screen.
The prior work also monitors a NIC's network traffic by a spy application that transmits from an NVMe SSD, reconstructing the victim's web browsing behavior from the intercepted traffic pattern.

\section{Proposed Approach}
\label{sec:approach}

\subsection{Motivations}
USB and PCIe utilize similar communication stacks and share bandwidth through hubs/switches in a similar manner. 
The logical extension of Invisible Probe would be to broaden the attack across different buses and to develop a more general use case. 
USB is the dominant interface for peripheral devices in everyday computing, especially with the prevalence of laptops among users~\cite{noauthor_idc_nodate} and the growing trend to condense many different ports to USB-C.
Meanwhile, the usage of PCIe devices is being reduced by the increased adoption of System-on-Chip (SoC) architectures (e.g. Apple M Series) in consumer devices, which integrate several peripheral controllers directly, further diminishing the role of PCIe in everyday computing.
Overwhelming the bandwidth of a southbridge or chipset through PCIe is also more challenging due to the much greater speed and bandwidth allocation of PCIe connections. 
USB's widespread use and relatively lower bandwidth make it a more vulnerable target for attacks.

We sought to adapt the principles used in Invisible Probe to situations specific to USB and develop new attacks. 
For a keystroke monitoring attack, as a GPU and specialized RDMA NIC are connected only through PCIe instead of USB hubs, we devised a novel strategy for recovering the timing information over USB hubs. 
Comparatively, our attacks on USB could be more widely executed with fewer restrictions.

\redtext{
Because USB operates as a shared bus with a fixed bandwidth available, when multiple devices attempt to communicate simultaneously at the maximum speed of the bus, their concurrent transfers create contention, leading to congestion and a reduction in the effective speed of each transfer. 
This congestion alters the timing and delays of data exchanges, which can be measured to infer information about the nature of other transfers. 
By analyzing these variations, it is possible to uncover patterns that reveal details about the data being transmitted, enabling side-channel attacks to exploit this shared-bus architecture. 
}
The primary objective was to establish and measure congestion scenarios over both USB 3.X and USB 2.0 root connections, as most modern USB devices with a USB 3.X or higher connector will default back to USB 2.0 speeds when connected to a lower-speed port.
Even though the bandwidth from USB 3.X to USB 2.0 is drastically reduced, the latency of a data transfer across both USB 3.X and 2.0 is not reduced, which helps our attack as it reduces any possible inherent differences between specifications that could make it less transferable between revisions~\cite{noauthor_usb_nodate-1}.

This section outlines the methodology used to create congestion scenarios, how the data was collected, and the machine learning methods to profile the gathered information. 
By carefully designing the experimental setup, we were able to simulate real-world conditions and capture relevant data. 
\redtext{This data was then analyzed using Hidden Markov Model (HMM) and Long Short-Term Memory (LSTM) machine-learning models to uncover patterns of keystrokes and websites, respectively,} demonstrating the feasibility and severity of such system-level side-channel attacks.

\redtext{
\subsection{Keystroke Attack Threat Model}
Previous research has shown the possibility of recovering characters typed into a computer based on the timing of their entry by a human user~\cite{song_timing_nodate}~\cite{zhang_peeping_2009}. 
While earlier studies exploited SSH network packets or system calls in multi-user systems, mitigations have since been proposed to suppress these side channels~\cite{song_timing_nodate}~\cite{schwarz_keydrown_2018}. 
However, this attack vector takes a new form by targeting the USB hub.
Traditional keyloggers, such as inline devices inserted between the keyboard and the USB port~\cite{noauthor_usb_nodate-2}, are specialized devices and require physical access, limiting their scalability. 
Our attack, in contrast, does not require additional devices or physical access and could be launched remotely at scale, making it significantly more dangerous. 
These attacks exploit the fundamental \textbf{Trust by Default} expectation that computers extend to USB devices. 
}

\redtext{
This attack leverages the inherent vulnerabilities of USB 2.0 hubs with a single TT.
Since USB keyboards utilize \textit{interrupt} data transfers instead of the more common bulk transfer, and the data packet size tends to be small, it is challenging for side-channels to have the resolution required to detect them. 
The spy device has to have similar transfer rates, USB version, and transfer type as the victim device, and a USB mouse satisfies all these requirements. 
An attacker could modify a USB device's firmware to emulate an additional USB mouse to the system and add spy functionality. 
To create congestion on the USB hub that an adversary can observe, the spoofed mouse device (mouse "jiggler") sends repeated, slight side-to-side movement commands. 
These commands are imperceptible to the user and do not interfere with standard mouse functionality.
As the victim keyboard sends data to the host and the spoofed mouse on the same USB hub sends commands to saturate the TT bandwidth, congestion arises and creates a side-channel. 
The spy device can measure the timing information for the mouse movement commands when the acknowledgment packet to a mouse movement is received, and this timing data will reveal the keystrokes.  
}

\subsection{Methodology for Recovering Keystroke Data}
For our experimental setup, a wired mouse~\cite{noauthor_keychron_nodate} and keyboard~\cite{noauthor_keychron_nodate-1} were connected to a USB hub. \redtext{We tested with different hubs (a USB 2.0 hub~\cite{noauthor_inland_nodate-2} and a USB 3.0 hub~\cite{noauthor_inland_nodate-1}) to validate that the attack works across USB hub generations. 
Each tested hub had a single TT.}
The test computer had multiple root hubs, which isolated the traffic to and from the two devices from the rest of the system. 
The mouse was locked in place and rigged to a continuously rotating surface, which gave the sensor on the mouse constant mouse movement updates. 
The usual hardware processed these updates and sent them to the host at the maximum polling rate. 
This created congestion on the TT so that each time the keyboard was used to input a character, there was a delay between mouse movement updates. 
In our setup, a spy program monitored the mouse input events, recorded the timings of each mouse update when each word was typed, and saved them in eight-second traces. 
This is also described through Algorithm \ref{alg:keyboardDelayCollection}.

The words were selected from a dictionary of 7658 common passwords between 4-10 letters in length, consisting of a 26-letter alphabet of English characters. 
We only recorded the words to the dataset if the typed sequence was correctly inputted and did not contain backspace keystrokes or capital letters. 

\begin{algorithm}[h]
\caption{Keystroke Timings Collection}
\label{alg:keyboardDelayCollection}
\begin{algorithmic}
\STATE Connect keyboard and mouse to Single-TT USB hub;
\STATE Lock mouse on rotating surface \& set max polling;
\STATE Open($word\_trace\_file$);
\STATE Initialize listener for mouse movement updates and key events to save events $word\_trace\_file$;
\FOR{$word \in$ CommonPasswords}
    \STATE RecordTimeSinceLastMouseUpdate();
    \STATE RecordTimeOfKeystrokes();
    \IF{!backspace \AND !capitals \AND $word$ is valid}
        \STATE SaveTrace($word\_trace\_file$);
    \ENDIF
\ENDFOR
\end{algorithmic}
\end{algorithm}

\redtext{
\subsection{Website Recovery Attack Threat Model}
Another facet of this research demonstrates an attack that infers website browsing behavior based on USB congestion patterns.
Research has shown that website traffic patterns can be deduced from their unique characteristics~\cite{yang_usb_2018}~\cite{gu_traffic-based_2019}. 
Our attack again targets congestion at the USB hub level with USB network adapters, but this attack can be executed on any speed class of USB hub.
An attacker accesses a USB device connected to the computer, e.g. a USB storage device, webcam, or other \textit{bulk transfer} device, and uses software to query the USB device.
The attacker then performs constant bulk data transfers to and from the device in a perceptibly normal and benign way, saturating the bandwidth of the shared hub.
When a USB network adapter (victim device) is connected alongside and used for website browsing, the spy device transfer latency is affected by the network activity.
These latency variations correlate with the network traffic required by different websites, creating an effective side-channel for fingerprinting websites.
}

\redtext{
An individual's browsing behavior is of significant privacy concern, as it can be a valuable commodity for advertising agencies, a tool for phishing attackers, and a target of interest in countries with strict rules on expression. 
The combination of website fingerprinting and keyboard input monitoring proposed in this work represents a formidable attack vector exposed by common USB hubs in the real world, presenting severe threats to the confidentiality and privacy of ordinary users.
}

\begin{figure*}[h]
    \centering
    \subfloat[\redtext{Spy devices  (red data) are set up to have full bandwidth of a USB hub}]{\includegraphics[width=0.49\textwidth]{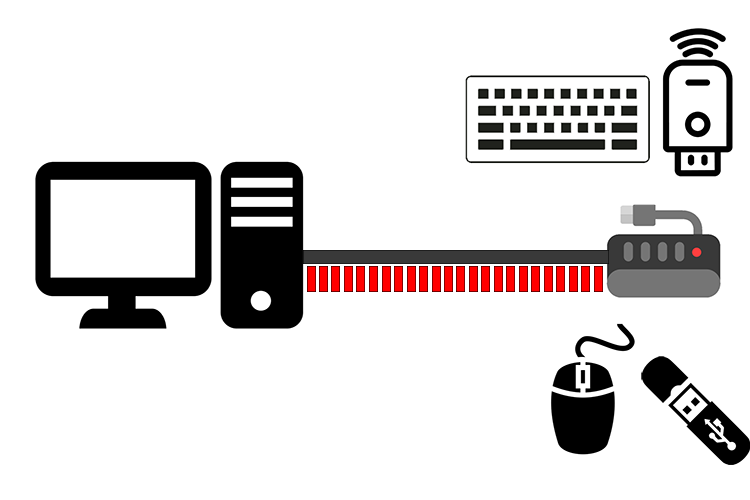}\label{fig:subfig1}}
    \hfill
    \subfloat[\redtext{At run-time, victim devices (blue data) share the bandwidth with the spy devices and  cause congestion}]{\includegraphics[width=0.49\textwidth]{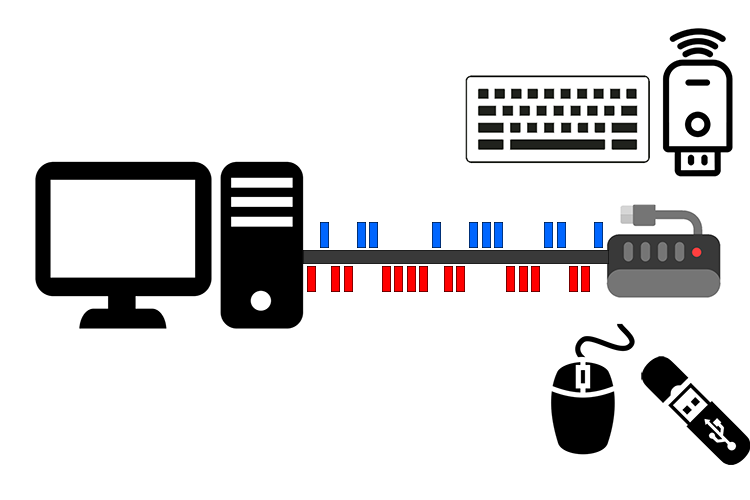}\label{fig:subfig2}}
    \caption{\redtext{Congestion on the shared bus and the side-channel observable by the spy device}} 
    \label{fig:congestion-example}
\end{figure*}

\subsection{Methodology for Recovering Web Traffic Information}
To create a congestion scenario to monitor web traffic from a USB ethernet adapter~\cite{noauthor_usb-c2500wired_nodate}, an external SSD~\cite{sabrent_usb_nodate} was selected as the spy device, and a large data file on it was read/transferred back to the host in batches. 
These devices were connected to the two previously listed hubs and a USB Type-C multi-port hub~\cite{noauthor_inland_nodate}, which also had a built-in ethernet adapter.
Our spy program initiated a continuous stream of 4 KiB data read operations from the SSD for approximately 8 seconds for each data trace to saturate the USB connection. 
The spy program used the \textit{io\_uring library} available in Linux, a high-performance asynchronous I/O framework. 
By leveraging \textit{io\_uring}, we could simultaneously queue multiple disk read requests to reduce the system overhead and measure the response time between each request.  
The key metric for detecting bus activity was the delay timing between disk reads, which increased with the bus's congestion level. 

One second after the spy program began its trace, a separate process launched the Google Chrome web browser in incognito mode with a cache-clearing directive. 
A selected website from a list of the top 100 websites in the US~\cite{hardwick_top_2023} was browsed, utilizing the USB Ethernet adapter. 
The side-channel data collection process is visually depicted in \autoref{fig:congestion-example}. 
The spy program measured the increased delay for its data transfer. As both the spy and the victim devices utilized \textit{bulk transfers} and thus had the same communication priority across the hub, the higher the USB traffic generated by the network adapter, the more congestion on the bus and, therefore, the longer delay the spy data transfer experienced. 
Algorithm \ref{alg:usbCongestion} describes these data collection steps.
The congestion measurement was repeated 150 times for each of the top 100 websites for the three USB hubs.
The dataset was then sanitized to ensure that congestion was measured across the bus to ensure that the web traffic was coming from the network and not a system cache. 
Traces with significantly shorter lengths or without any deviation from the average delay were identified as erroneous, primarily due to failed web page loading or occasional USB dropouts. 
This step was crucial to ensure the quality and reliability of the training data, as such anomalies could skew the model's learning process and affect its accuracy.  
The data was input into a machine-learning model to profile the websites browsed. 

\begin{algorithm}[htb]
\caption{USB Web Traffic Congestion Data Collection}
\label{alg:usbCongestion}
\begin{algorithmic}
\ENSURE USB data is being transferred by the USB NIC
\STATE Initialize SSD with Large Data File; 
\STATE Connect devices to the hub;
\STATE Initialize $io\_uring$ queues;
\STATE $top\_websites\_list = ReadTopWebsitesFile()$;
\FOR{$i \leftarrow 1$ \TO $150$}
\FOR{$website \in top\_website\_list$}
    \STATE $previous\_time \leftarrow \text{GetCurrentTime()}$;
    \STATE Initiate all 4 KiB Read Requests;
    \STATE $requests\_to\_complete \leftarrow 81920$;
    \STATE $timeline[] \leftarrow \text{NewList()}$;
    \STATE BrowseToWebsite($website$);
    \WHILE{$requests\_to\_complete > 0$}
        \STATE WaitForReadRequestResponse();
        \STATE $current\_time \leftarrow \text{GetCurrentTime()}$;
        \STATE $timeline[requests\_to\_complete] \leftarrow current\_time - previous\_time$;
        \STATE $previous\_time \leftarrow current\_time$;
        \STATE RemoveCompletedRequestFromQueue();
        \STATE $requests\_to\_complete \leftarrow requests\_to\_complete - 1$;
    \ENDWHILE
    \STATE SaveToFile($timeline$, $website$);
\ENDFOR
\ENDFOR
\end{algorithmic}
\end{algorithm}

\section{Experimental Results of Attack 1 - Keystroke Recovery}
\label{sec:results1}

\subsection{Keystroke Timing Accuracy}

\begin{figure}[b]
    \centering
    \includegraphics[width=\linewidth]{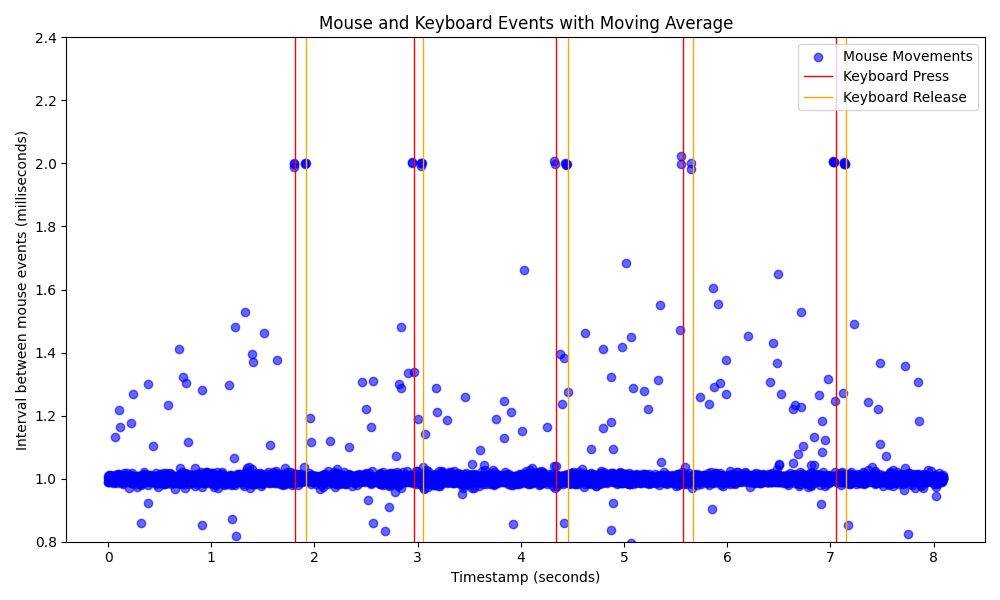}
    \caption{Collected mouse latency measurements and key presses}
    \label{fig:mouse}
\end{figure}

To determine the feasibility of keystroke recovery through the USB congestion-based side-channel, we first need to determine whether the keys being typed could be identified based on the timing patterns of mouse movements captured via the side-channel. 

\redtext{
A typical USB mouse can input commands every one millisecond.
When measuring the mouse updates in our test setup, we observed that most updates are delivered within one millisecond, indicating no congestion. 
Some updates have noticeably longer delays, indicating possible bus congestion within the TT due to keystrokes. 
With the chosen keyboard for experiments, a packet of data is sent to the host each time a key is pressed or released, causing congestion on the TT. 
This is visualized in \autoref{fig:mouse} with the samples of 2-millisecond intervals. 
We also monitored the timings of each keystroke event (the press and release) as a direct channel, shown as the two types of vertical lines in \autoref{fig:mouse}, to verify that the delayed mouse updates are closely aligned with the keyboard press and release events.
After filtering out single instances above a 1.8-millisecond threshold and all data points below that threshold, we could determine when key presses occurred very consistently. 
}

\begin{figure}[h]
    \centering
    \includegraphics[width=\linewidth]{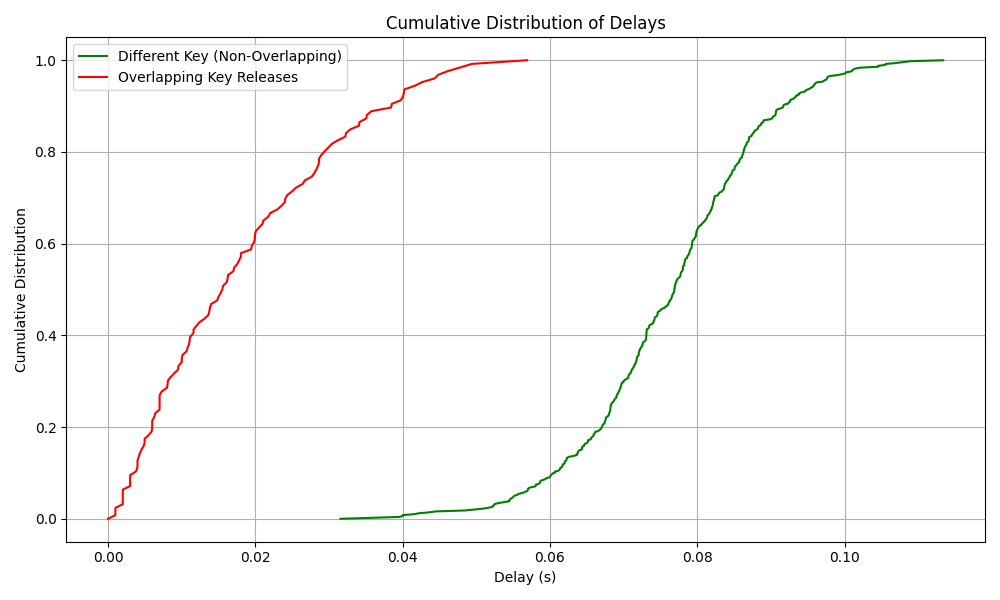}
    \caption{Cumulative distribution curves for delays between overlapping and distinguished keystrokes}
    \label{fig:overlap}
\end{figure}

A complication we encountered when recording one of our participants' typing behavior for the HMM model was that some keystrokes would overlap, meaning they began the next key press before the previous key was released, so some updates might be incorrectly filtered. 
When we analyzed how often this occurred, we found that approximately up to 10.4\% of their keystroke traces included an overlapping keystroke. 
This prompted a further analysis of the data set to understand how this would affect our results. 
After inspecting the behavior directly, we discovered that the vast majority of keystrokes that overlap have events that are timed much closer together than typical sequential keystrokes, which is visualized in \autoref{fig:overlap}. 
We set a threshold of 50 milliseconds between keystroke events to filter when overlapping keystrokes occur, and we were able to distinguish and label 98\% of keystroke events correctly and found no noticeable depreciation in the following HMM profiling results even with a few incorrectly labeled key press timings. 
The specific hub did not affect these results, as even when connected to a USB 3.X hub, USB mice and keyboards will still use the TT built into the separate USB 2.0 data path within the hub.

\begin{table*} 
    \centering
    \caption{Password Recovery Accuracies}
    \label{tab:key_model}
    \begin{tabular}{|l|l|r|r|} 
        \hline
        \textbf{Dataset} & \textbf{Side-channel} & \textbf{Top-10 Accuracy} & \textbf{Top-50 Accuracy} \\ 
        \hline
        \hline
         7658 Words, 26 Letters &USB Keyboard and Mouse side-channel (Our work) & 36.3\% & 89.3\% \\
        \hline
\multirow{2}{*}{1000 Words, 10 Letters} & Invisible Probe~\cite{tan_invisible_2021}  & 66.7\% & 95.3\% \\
        \cline{2-4}
       &  USB Keyboard and Mouse side-channel (Our work) & 69.2\% & 96.4\% \\
        \hline
      \multirow{2}{*}{4500 Words, 15 Letters} &  Peeping Tom in the Neighborhood~\cite{zhang_peeping_2009} & 38.0\% & 86.0\% \\
        \cline{2-4}
       & USB Keyboard and Mouse side-channel (our work) & 55.8\% & 93.2\% \\
        \hline
    \end{tabular}
\end{table*}

\subsection{Keystroke Recovery via Timing Information}
\begin{figure}
    \centering
    \includegraphics[width=0.7\linewidth]{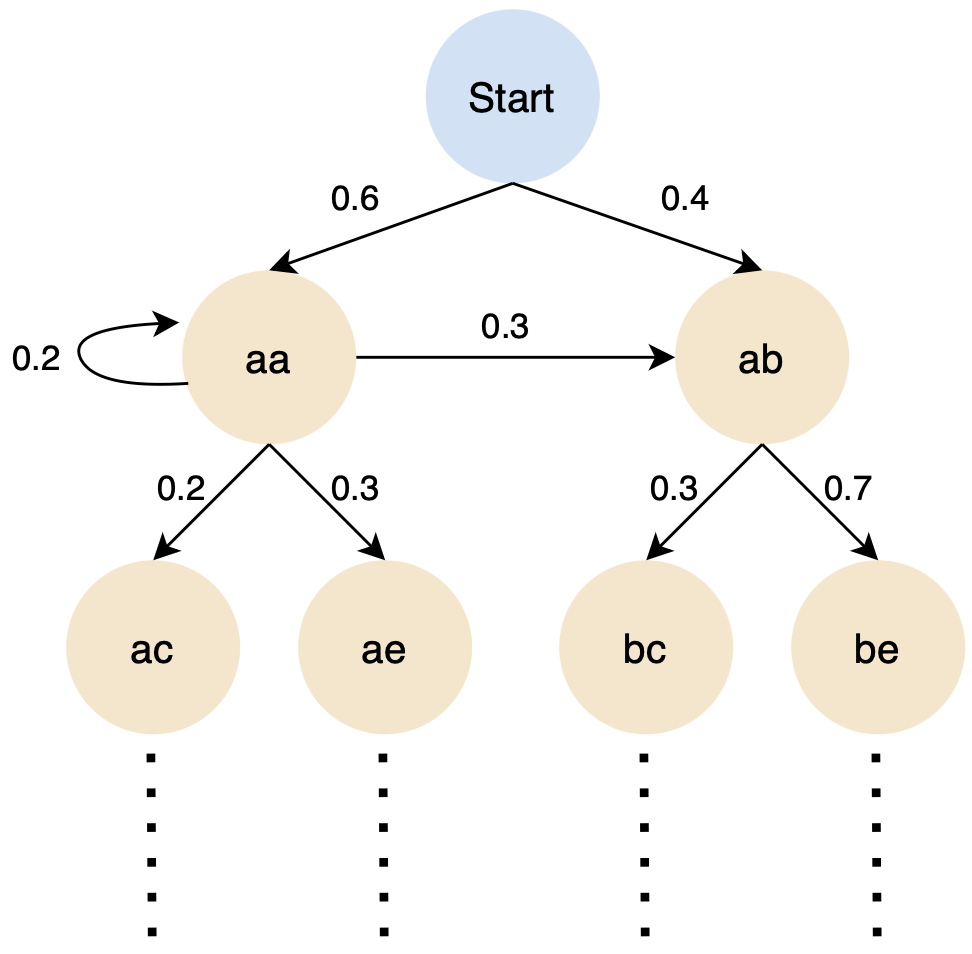}
    \caption{Simplified visualization of the HMM. \redtext{It processes sequential keystroke timings and determines the most likely sequence of pairs/characters matching the observed timing information. Each edge is a probability of transitioning from one state to another.}}
    \label{fig:hmm}
\end{figure}

We focused on the recovery of passwords - a sequence of characters derived from the delay between keystrokes. 
We used the most typical subset of 7658 passwords from a common password library~\cite{password}, using all 26 letters in the English alphabet. 
The designed HMM, as depicted in \autoref{fig:hmm}, consisted of hidden states and transitions between those states. 
Each state is a pair of characters, and the latency between the two keystrokes of the character pair is the output observation from the character-pair state. 
\redtext{
An HMM model was built from the profiled keystrokes from the USB side-channel. 
A timing trace collected at runtime while a password was typed is fed to the HMM model for sequence prediction. 
The core of our analysis involved the $n$-Viterbi algorithm. 
Unlike the naive Viterbi that only predicts the most likely sequence, the $n$-Viterbi algorithm returns $n$ most probable sequences.
} 

\redtext{
We report both the Top-10 accuracy (the correct word was within the top 10 highest prediction probabilities) and Top-50 accuracy (the correct word was within the top 50 predictions).
}
We achieve a \textbf{top-10 accuracy} of 36.3\% and a \textbf{top-50 accuracy} of 89.3\%. 
Due to a difference in datasets and recovered alphabet, these results are not directly comparable to the previous work~\cite{tan_invisible_2021, zhang_peeping_2009}.
Invisible Probe~\cite{tan_invisible_2021} built their model with 1000 words and a 10-letter alphabet (``etaoinshrd"). 
Peeping Tom in the Neighborhood~\cite{zhang_peeping_2009} utilized a dataset of 4500 words and a 15-letter alphabet.
To build direct comparisons to these works, we repeated the attack with the same sized datasets as the prior work, and the results are shown in \autoref{tab:key_model}, demonstrating that our side-channel achieves better keystroke recovery under the same dataset. 
It can be seen that the accuracy of the models increases by constraining the dataset to fewer letters and words. 
However, even a full 26-key dataset still has a high enough accuracy to be potentially dangerous in the real world. 
\redtext{Combined with a language analysis on the predicted words, an attacker could further improve the recovery of the typed words with fewer guesses.}
With further investigation, better algorithms than $n$-Viterbi and HMM may allow for higher accuracies and transferability so that the models are not specific to one person's typing patterns.

\section{Experimental Results of Attack 2 - Website Fingerprinting}
\label{sec:results2}

\begin{figure*}
    \centering
    \hfill
    \subfloat{\includegraphics[width=0.49\textwidth]{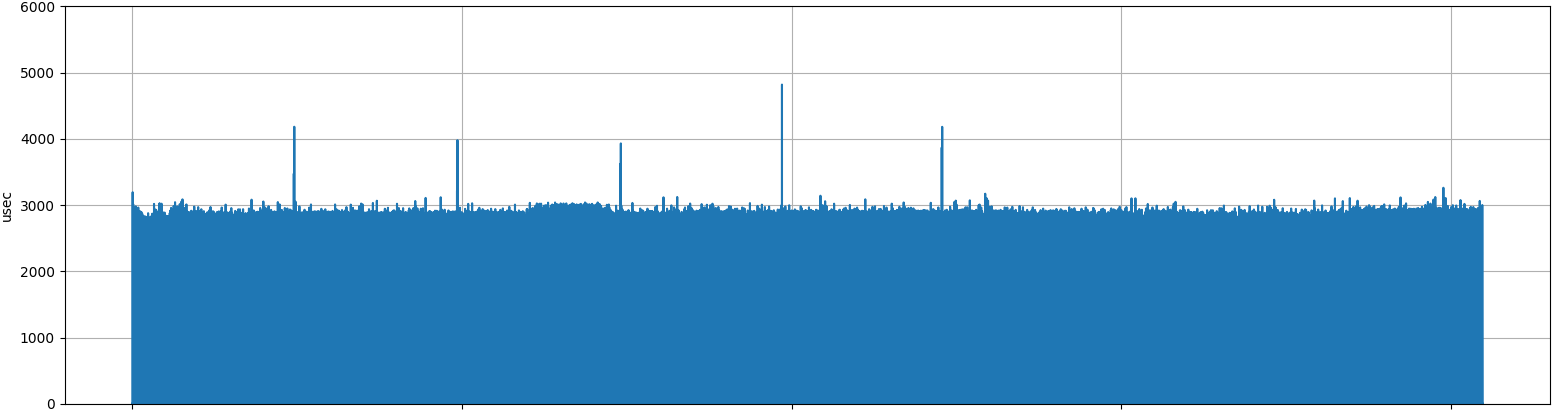}}
    \hfill
    \subfloat{\includegraphics[width=0.49\textwidth]{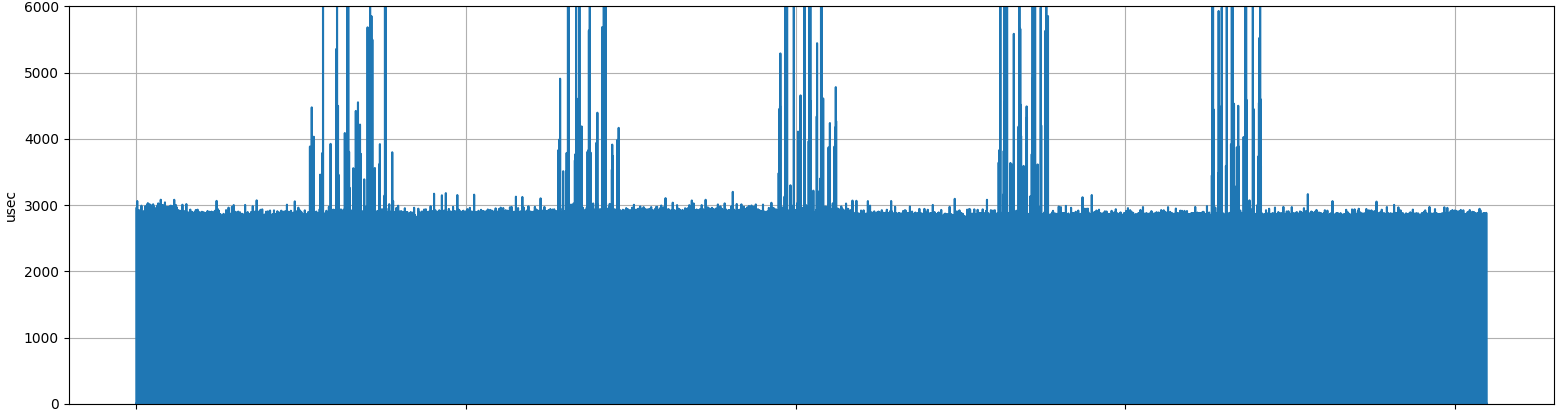}}
    \caption{\redtext{Timeline plots of the measured USB disk congestion detection with 128 KiB and 4 MiB bursts of arbitrary network traffic}}
    \label{fig:resolution}
\end{figure*}

\subsection{Resolution of USB Disk/Bulk Transfer Spy}
\label{subsec:resolution}

This phase of the experiment aimed to determine the minimum network traffic volume needed to create discernible congestion on the USB bus, as detected by the disk reading spy mechanism. 
This determination is crucial for understanding the sensitivity and practical implications of USB congestion-based side-channel attacks.
A local server was set up to receive network traffic, and a process generated controlled bursts of random data packets ranging from 16 B to 4 MiB. 
Each burst was transmitted five times with a one-second delay between transmissions, allowing systematic analysis of traffic size and resultant USB congestion.

Our spy mechanism observed some congestion at 64 KiB data sizes, though inconsistently, while at 128 KiB, the disk reading spy consistently detected congestion, confirming this as a reliable threshold for observable USB bus congestion. 
This is shown in \autoref{fig:resolution}, alongside 4 MiB bursts of arbitrary data to demonstrate unambiguous congestion.
These findings provide a benchmark for the traffic volume necessary to induce USB congestion that is exploitable reliably in side-channel attacks. 
While most individual HTML, CSS, and Javascript files on the internet are smaller than this threshold, the cumulative traffic from concurrent file transfers, images, and videos typically exceeds 128 KiB, making this benchmark valuable for further research and protective measures against similar USB-based attacks.

\subsection{Congestion Correlation with Direct Network Data}
To validate that the congestion observed on the USB bus when an SSD disk spied on an ethernet adapter was directly attributable to the volume of network traffic, we conducted a comprehensive analysis to establish a definitive correlation between the network traffic volume captured by the network adapter (victim direct channel) and the congestion data recorded by our monitoring mechanism (via the side-channel). 
We used Wireshark~\cite{wireshark} (the network protocol analyzer) to capture the traffic received by the USB network adapter, providing a direct and unambiguous channel for examining the volume of network data that traversed the USB NIC. 
We then calculated the correlation between the volume of traffic measured by the Wireshark process and the maximum delay experienced by the USB disk spy. 

\begin{figure}[b]
    \centering
    \includegraphics[width=\linewidth]{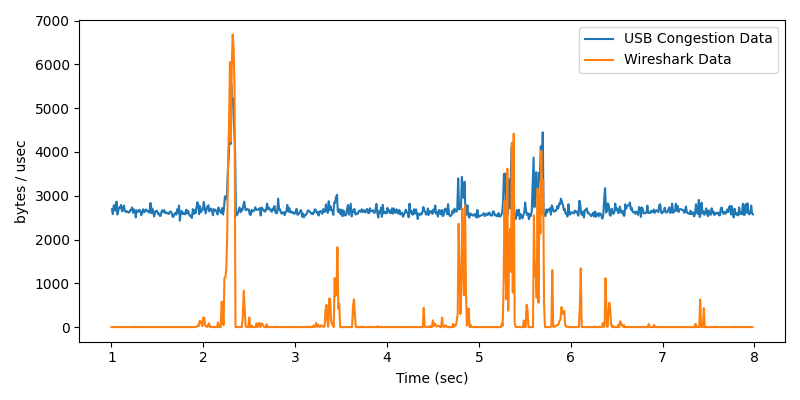}
    \caption{Effectiveness of the USB Congestion Side-Channel Data Correlating with the Wireshark Direct-Channel Data}
    \label{fig:correlation-graphs}
\end{figure} 

The correlation calculation results were reassuringly positive, and an example can be viewed in \autoref{fig:correlation-graphs}. 
The orange line is the direct-channel trace, where each datapoint is the summed total volume of Wireshark network traffic in bytes in a 5 ms window. 
The blue line depicts our side-channel congestion measurement, which is the maximum latency of the disk spy in milliseconds within the same 5 ms window. 
For almost every website tested, the correlation coefficient between the orange and blue lines exceeded 0.5, \redtext{often reaching a coefficient of 0.8-0.9}. This high level of correlation strongly supports the hypothesis that the congestion observed on the USB bus was indeed a direct consequence of the network traffic volume. 
The findings from this Wireshark analysis and subsequent correlation calculation provide robust validation for our methodology.

\subsection{Supervised Website Fingerprinting via Congestion Side-channel}

\begin{figure}[t]
    \includegraphics[width=\linewidth]{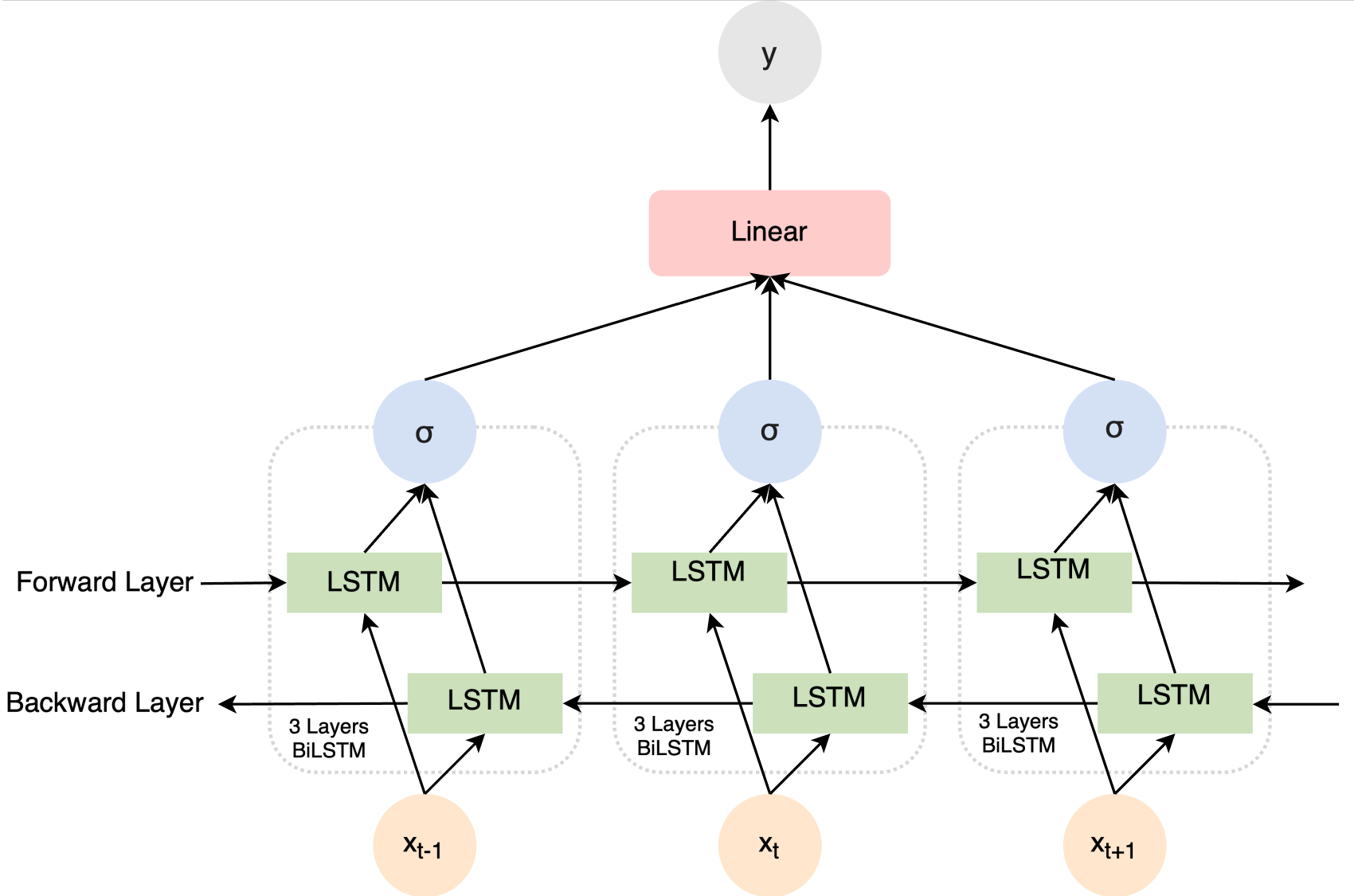}
    \caption{Example BiLSTM model. \redtext{For website profiling, it can learn backward and forward dependencies of time-series/sequential data, which is designed to improve the model's accuracy in predicting web pages from congestion traces.}}
    \label{fig:lstm}
\end{figure}

We employed a Bidirectional LSTM (BiLSTM) model~\cite{bilstm}, which is a form of recurrent neural network (RNN) that can capture dependencies in both forward and backward directions of time-series/sequential data. \autoref{fig:lstm} depicts an example model. 
The bidirectional nature of the LSTM units enabled the model to capture temporal dependencies in both directions, enhancing its ability to discern subtle variations in the congestion patterns that could indicate specific web browsing behaviors. 
This bidirectional approach is particularly effective in scenarios where the context of the input sequence is crucial for making predictions. 
The BiLSTM model processed the traces through several layers, beginning with the LSTM layers that learned from both forward and backward dependencies in the data. 
This was followed by a fully connected layer that translated the LSTM outputs into predictions for website classification. 
To validate the model's performance, we employed a 5-fold cross-validation approach.

\begin{table} 
    \centering
    \caption{ML Model Accuracies for Website Classifier}
    \label{tab:model_accuracies}
    \begin{tabular}{|l|r|r|} 
        \hline
        \multirow{2}{*}{\textbf{Model}} & \multicolumn{2}{|c|}{\textbf{Accuracy}}  \\ 
        \cline{2-3}
         & \textbf{Top-1} & \textbf{Top-3} \\  
         \hline
         \hline
        USB 2.0 Hub on Trained Network & 83.4\% & 89.2\% \\
        \hline
        USB 3.X Hub on Trained Network & 81.1\% & 88.9\% \\
        \hline
        USB-C Hub on Trained Network & 80.6\% & 87.9\% \\
        \hline\hline
        USB 2.0 Hub on Untrained Network & 78.2\% & 84.7\% \\
        \hline
        USB 2.0 Hub on Untrained VPN Network & 70.7\% & 78.2\% \\
        \hline
        USB 2.0 Hub on Trained VPN Network & 81.1\% & 87.9\% \\
        \hline
    \end{tabular}
\end{table}

Our website fingerprinting results, summarized in \autoref{tab:model_accuracies}, demonstrate the performance of the BiLSTM model under various scenarios. 
With the same set of victim and spy USB devices previously outlined, we first collected traces from an office computer in a commercial network. 
We report both Top-1 accuracy (the correct website has the highest prediction probability) and Top-3 accuracy (the correct website was within the top three predictions).
Using the USB 2.0 hub, the model achieved a \textbf{Top-1 accuracy} of 83.4\% and a \textbf{Top-3 accuracy} of 89.2\%.
The accuracies for USB 3.X hubs are similar. 
The results are shown in the first two rows of \autoref{tab:model_accuracies}.
We further evaluated the model with a USB Type-C multi-port hub, where the spy device is the same external SSD, but the victim device is the integrated ethernet adapter.
The accuracies for the USB-C hub are also similar, indicating that the congestion-based side-channel is not hub-specific and should work across all different versions of USB hubs
It also indicates that the attack is not device-specific.

To assess the transferability of the model for different network conditions, we set up the same USB 2.0 hub with the same set of devices on a different computer, which was on a physically different residential network. 
We collected the congestion data and directly applied the model trained for the office setup, which achieved a Top-1 accuracy of 78.2\% and a Top-3 accuracy of 84.7\%, as shown in the fourth row of \autoref{tab:model_accuracies}.
This result demonstrates the robustness of the model to varied network conditions. 
Additionally, we tested the pre-trained model on another dataset collected from the office computer connected through a Virtual Private Network (VPN), obtaining a Top-1 accuracy of 70.7\% and a Top-3 accuracy of 78.2\%, shown in the fifth row of \autoref{tab:model_accuracies}. 
These results were lower without retraining, given the encryption overhead and additional delay introduced by the VPN affecting the website traffic.
When we combined the VPN dataset with the initial USB 2.0 dataset into a comprehensive `super-set' and retrained the model, this combined training resulted in a Top-1 accuracy of 81.0\% and a Top-3 accuracy of 87.9\%, as indicated on the sixth row of \autoref{tab:model_accuracies}. 
These results underscore the model's ability to generalize across different network environments and conditions, highlighting the versatility and transferability of this USB congestion-based side-channel attack. 
Importantly, even when users employ VPNs for privacy, their browsing behaviors remain vulnerable to this side-channel attack, exposing private internet activity through USB traffic.

By comparison, the previous Invisible Probe work reported a Top-1 accuracy of 87.4\% and a Top-3 accuracy of 96.1\% for their BiLSTM model~\cite{tan_invisible_2021}. 
Their AttLSTM model achieved higher accuracies, with a Top-1 accuracy of 93.3\% and a Top-3 accuracy of 98.7\%~\cite{tan_invisible_2021}. 
While our results are slightly lower than those of the prior work based on PCIe, the lower latency and greater bandwidth of PCIe compared to USB possibly result in a higher signal-to-noise ratio of the side-channel and slightly more discernible data patterns.
Also, because the environment where we collected data is in residential and commercial networks instead of a data center, our results paint a more accurate picture of the network environment that typical users encounter, thus validating the applicability of this attack.

\section{Proposed Countermeasures and Mitigation}
\label{sec:mitigation}
\redtext{
The USB Implementers Forum (USB-IF) has not standardized specific protections against these types of attacks, although some countermeasures under consideration include encrypting USB data payloads between the device and host. 
While encrypting USB packet payloads could theoretically prevent inline snooping devices from directly reading the data, it would not mitigate a timing attack against the keyboard. 
The timing patterns of human keyboard input would still persist, rendering our timing attack effective even with encrypted USB data packets.
To deploy the keystroke attack against many users, the devices with emulated mice could either be sold through retail channels directly, hacked into reputable devices, or put in USB drives that are left around public and workplaces for anyone to take and use~\cite{nohl_badusb_2014}~\cite{babu_what_2022}~\cite{tian_sok_2018}.
A mouse device normally is not a threat because a mouse does not receive positional feedback about where it is on the screen to create malicious interactions.
}

One straightforward mitigation strategy users can adopt to protect against USB congestion-based side-channel attacks is the proactive management of their device connections. 
Users can reduce the number of active connections by simply unplugging peripheral devices that are not in use, thereby minimizing the potential for USB congestion and reducing the attack surface. 
Additionally, USB hubs with physical switches offer a more conveniently controlled environment. 
These switches allow users to manually enable or disable specific ports, cutting off unnecessary data paths and reducing opportunities for unauthorized data capture via congestion.
Although some other mitigations exist, such as requiring user authorization for USB connections, the most common advice for USB security is to avoid plugging unknown devices into personal computers. 
Consequently, this attack is becoming less effective, although it remains a prevalent threat~\cite{tian_sok_2018} for many users.

More proactive users can also use specialized security-focused operating systems such as Quebs OS~\cite{noauthor_qubes_nodate} to remove the \textit{Trust by Default} of USB, but this requires a much more drastic change for many users and will reduce the usability of a system as USB keyboards and mice must manually be given permission to interact with the system. 
This will mitigate the BadUSB-style USB mouse spy against the keyboard as that attack introduces a new device to the system. 
It can also isolate USB devices from unknown programs by preventing access to the USB Qube. 
This mitigation strategy most likely would need similar implementations in Windows and MacOS to be widely accepted.

Modifying how USB drivers handle concurrent device traffic at the software level can also provide protection. 
USB driver designers could develop mechanisms to randomize bandwidth allocation to connected devices, complicating an attacker's ability to predict or influence traffic patterns and thus acquire reliable data from congestion, similar to what~\cite{schwarz_keydrown_2018} recommends. 
Moreover, altering the USB traffic arbitration algorithm to remove fairness guarantees could also serve as a deterrent. 
Making traffic less predictable and not necessarily fair from a device's perspective makes it challenging for an attacker to manipulate or infer data based on congestion patterns.

Changes to how motherboards and computers are structured on the hardware design front can play a crucial role in mitigating these attacks. 
Designing motherboards and systems with multiple USB root hubs can effectively split traffic among different controllers, preventing congestion by distributing the data load across the system. 
This makes it significantly more challenging for a congestion-based side-channel attack to succeed.

\redtext{Though this research only profiled the top 100 websites in the US, real attacks could scale up to 1000s of websites to be more comprehensive. They could also specifically target certain websites, like banking or financial websites, to identify when the user is browsing a website with highly sensitive/coveted data and deploy further attacks on the system.} 

Protecting against USB congestion-based spy attacks requires a multi-faceted approach, including user vigilance, software enhancements, and hardware design improvements. 
While simple measures like unplugging unused devices and using switchable hubs provide immediate, user-driven solutions, changes in USB driver behavior and system architecture will likely result in more substantial protection. 
By addressing the issue at multiple layers, from the user to the hardware design, the security of USB connections can be significantly enhanced, safeguarding data against sophisticated side-channel attacks.

\section{Future Research}
\label{sec:extensions}
\subsection{Other USB Devices}
The congestion-based side-channel can theoretically be used to spy on other USB devices, such as USB-based security keys and USB mice. 
As more websites support USB-based security keys for authentication, it is crucial to understand their vulnerabilities to previously unconsidered congestion-based side-channels. 
Investigating these vulnerabilities could provide valuable insights into enhancing the security of authentication mechanisms.

Additionally, analyzing USB mouse data could offer an understanding of user behavior or detect subtle user interactions that might reveal sensitive information. 
Beyond USB mice, desktop touch interface devices often connect through USB, making them potential targets for attacks that could disclose sensitive information valuable in other contexts. 

\subsection{Other Mediums and Shared Buses}
Another intriguing direction involves applying congestion-based side-channel techniques to Wi-Fi networks and other RF communication methods. 
By remotely probing the RF fields for activity, it may be possible to infer data transmission patterns or even specific activities occurring over the Wi-Fi network without even having to be connected to or have access to the physical machine.

If these remote Wi-Fi attacks can achieve similar website profiling performance as demonstrated in direct attacks on USB hubs of the previous section, it would necessitate enhanced protection measures for Wi-Fi. This would significantly expand the scope of side-channel analysis considerations in the design of wireless communication mediums.

\subsection{Using USB as a CPU Architectural Side-Channel}
\begin{figure}
    \includegraphics[width=0.9\linewidth]{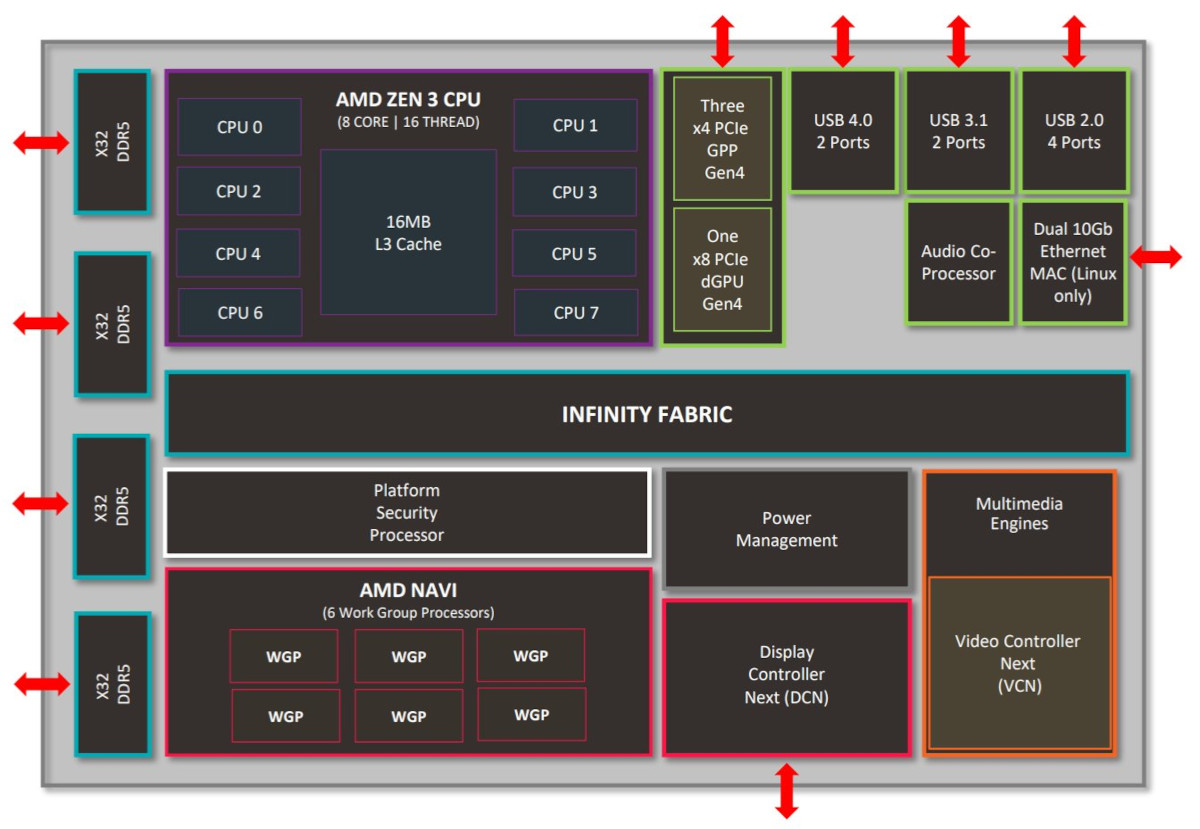}
    \caption{AMD Zen 3 APU Block Diagram, from~\cite{noauthor_amds_nodate}}
    \label{fig:AMDSOC}
\end{figure}

A promising area of research also lies in exploring USB-based side-channels within modern processors, especially in systems where the USB controller is integrated into the SoC, such as Apple's M series and AMD's Ryzen series CPUs (e.g. the core shown in \autoref{fig:AMDSOC}). 
The integration of USB controllers in these advanced SoCs presents a unique opportunity to study side-channel vulnerabilities at a more fundamental level of the processor architecture. 
Investigating USB-based side-channels in these processors could be as valuable as studying cache or ring-bus-based side-channels. 
The embedded nature of USB controllers in SoCs may reveal new insights into processor-level data flows and potential security vulnerabilities.

Exploration of congestion-based side-channels in modern SoCs, applying these techniques to wireless mediums like Wi-Fi, and extending various USB devices represent significant and uncharted territories in cybersecurity research. 
These studies have the potential to unveil novel insights into processor architecture vulnerabilities, wireless communication security, and peripheral device integrity. 
The ultimate goal of this research was to deepen our understanding of these potential security threats and develop robust countermeasures to protect against them.

\section{Conclusion}
\label{sec:conclusion}
This research has shed light on the intricate vulnerabilities associated with USB security, mainly through the exploration of congestion-based side-channel attacks. 
Our findings reveal the feasibility of such attacks and their potential sophistication, as evidenced by the substantial accuracy achieved in recovering keyboard inputs and profiling web browsing behavior. 
The successful adaptation of these techniques in varied network environments underscores the broader implications of our work. 
In the future, we plan to establish direct contact with the USB-IF so as to collaborate on developing solutions to mitigate these vulnerabilities. 
The expansion of this research to explore similar vulnerabilities in other mediums, such as Wi-Fi, opens new frontiers in our pursuit of robust cybersecurity defenses. 
As we continue to unravel the complexities of congestion-based side-channels, we aim to ensure a secure and resilient technological ecosystem.

\subsection{Acknowledgements}
This work was supported in part by National Science Foundation under grants IUCRC-1916762 and CNS-2212010.

\newpage

\input{ieee-main.bbl}

\pagenumbering{Arabic}

\end{document}